\documentstyle[fleqn,prl,aps,preprint]{revtex}

\begin{document}
\draft
%\twocolumn
\widetext

\title{Dynamics of coupled quantum spin chains}
\author{H.J. Schulz}
\address{
Laboratoire de Physique des Solides,
Universit\'{e} Paris--Sud,
91405 Orsay,
France }
\maketitle

\begin{abstract}
Static and dynamical properties of weakly coupled antiferromagnetic spin
chains are treated using a mean--field approximation for the interchain
coupling and exact results for the resulting effective one--dimensional
problem. Results for staggered magnetization, N\'eel temperature and spin
wave excitations are in agreement with experiments on $\rm KCuF_3$. The
existence of a narrow longitudinal mode is predicted.  The results are in
agreement with general scaling arguments, contrary to spin wave theory.
\end{abstract}
\pacs{75.10Jm, 75.30Cr, 75.30Ds, 75.50Ee}

\narrowtext 

One--dimensional quantum spin chains are interesting objects to study for a
number of reasons. On the one hand, experimental systems are generally very
well described by simple yet nontrivial Hamiltonians involving very few
unknown parameters. The standard example is the Heisenberg model, with only
one free parameter, the exchange constant. Comparison between theory and
experiment then becomes a particularly stringent test, as a variety of data
have to be explained by one single parameter. Moreover, both a large number
of exact theoretical results and powerful analytical and numerical methods
are available, making this comparison particularly interesting. On the other
hand, in spite of their simplicity, models of quantum spin chains have lead
to a number of unexpected and unconventional predictions. For example, the
exact solution of the antiferromagnetic spin--1/2 Heisenberg chain shows
that the low--lying excitations are spin--1/2 objects \cite{bethe_xxx} (now
called spinons), quite different from standard spin waves. This prediction
has been confirmed experimentally quite recently in $\rm
KCuF_3$.\cite{tennant_kcuf} Another example is Haldane's prediction of a gap
in the excitation spectrum for integer--$S$ antiferromagnets
\cite{haldane_gap} which again has found experimental
confirmation.\cite{buyers_csnicl,renard_nenp}

Strictly one--dimensional models of course do not exhibit phase transitions
into states with a broken symmetry. It is nevertheless clear that any real
compound, like $\rm KCuF_3$,
\cite{tennant_kcuf,satija_kcuf,tennant_kcuf_q1d}, $\rm
Sr_2CuO_3$, \cite{keren_srcuo,ami_srcuo} or $\rm Yb_4As_3$ \cite{fulde_ybas}
some form of interchain coupling is present. Then three-dimensional magnetic
long--range order can appear below a N\'eel temperature $T_N$. In the
present paper I show that in this case a conceptually simple approach,
namely treating the interchain coupling in the mean--field approximation and
treating the resulting effective one--dimensional problem as exactly as
possible, \cite{scalapino_q1d,schulz_gl_q1d} gives a coherent description
of the ordered state and produces nontrivial quantitative predictions for
static and dynamic quantities that can be successfully compared to
experiments, in particular on $\rm KCuF_3$ where detailed neutron scattering
results are available.\cite{tennant_kcuf,satija_kcuf}

I start with the natural model for coupled spin--1/2 antiferromagnetic
chains, namely a spatially anisotropic Heisenberg model for parallel chains
forming a square lattice (i.e. the lattice has tetragonal symmetry):
\begin{equation} \label{eq:h0}
H = J \sum_{i,\bbox{r}} \bbox{S}_{i,\bbox{r}} \cdot \bbox{S}_{i+1,\bbox{r}} 
+ J_\perp \sum_{i,\bbox{r},\bbox{\delta}} 
\bbox{S}_{i,\bbox{r}} \cdot \bbox{S}_{i,\bbox{r}+\bbox{\delta}} \; .
\end{equation}
Here $i$ and $\bbox{r}$ label lattice sites along the  chain ($z$) and
perpendicular ($x$, $y$) directions, $\bbox{\delta}$ is summed over the two
nearest neighbor vectors in the transverse directions, and
$\bbox{S}_{i,\bbox{r}}$ is a spin--1/2 operator at lattice site
$(i,\bbox{r})$. The longitudinal and transverse exchange constants are $J$
($>0$) and $J_\perp<0$, where in order to be close to the situation in $\rm
KCuF_3$ a ferromagnetic interchain coupling is used (but all of the
subsequent results apply with minor modifications also to $J_\perp>0$).

I now treat antiferromagnetic order using a mean--field treatment of the
interchain coupling. Assuming the order to be oriented along the
$z$--direction in spin space, the Hamiltonian (\ref{eq:h0}) transforms into
an effective single--chain problem described by \begin{equation}
\label{eq:h1} H_1 = J \sum_i \bbox{S}_{i} \cdot \bbox{S}_{i+1} -h \sum_{i}
(-1)^i S^z_i -2NJ_\perp m_0^2 \; .  \end{equation} Here $N$ is the number of
sites in the chain, $m_0 = (-1)^i \langle S^z_i \rangle$ is the staggered
magnetization, and $h=-4J_\perp m_0$. We thus have a one--dimensional
antiferromagnet in an effective staggered field $h$, with the order
parameter $m_0$ to be determined by minimizing the energy. The next step is
to transform (\ref{eq:h1}) into a fermionic model by a Jordan--Wigner
transformation and to go to the continuum limit. The resulting fermionic
model then is described by \begin{equation} \label{eq:h2} H_2 = \int dz \,
\left\{ -iv (\psi^\dagger_L \partial_x \psi^{\phantom{\dagger}}_L -
\psi^\dagger_R \partial_x \psi^{\phantom{\dagger}}_R) - h (\psi^\dagger_L
\psi^{\phantom{\dagger}}_R + \psi^\dagger_R \psi^{\phantom{\dagger}}_L) + 2g
\psi^\dagger_L\psi^\dagger_R \psi^{\phantom{\dagger}}_R
\psi^{\phantom{\dagger}}_L \right\} -2NJ_\perp m_0^2 \; .  \end{equation}
Here $\psi_{L,R}$ are standard fermion field operators for left-- and
right--moving fermions. Eq.(\ref{eq:h2}) can be recognized as the massive
Thirring model for which an exact Bethe ansatz solution
exists.\cite{bergknoff_mtm_exact} The parameter $v$ determines the velocity
of the low--lying excitations, and by comparing with the exactly known (for
$h=0$) spectrum of $H_1$ can be fixed as $v=\pi J a/2$ where $a$ is the
lattice constant. The proper identification of $g$ is more delicate because
the underlying fermionic lattice model has rather strong interaction and a
naive transition to the continuum limit therefore is uncontrolled. One can
however notice that the exact solution of $H_2$ has a mass gap
\cite{bergknoff_mtm_exact} \begin{equation} \label{eq:m} \Delta = h
e^{\Lambda(1-\gamma)} \frac{\tan \pi\gamma}{\pi(\gamma-1)} \; ,
\end{equation} where $\pi/\gamma = 2 {\rm arccot} (-g/2v)$ parameterizes the
interaction, and $\Lambda$ is the ``rapidity
cutoff''.\cite{bergknoff_mtm_exact} Requiring further that the total
particle number is independent of $m_0$, one has $\Lambda \propto \ln
(1/h)$, i.e. $\Delta \propto h^\gamma$. On the other hand, standard scaling
relations for the lattice Hamiltonian $H_1$ imply a mass gap $\propto
h^{2/3}$, i.e. we have $\gamma=2/3$. Thus the parameters in $H_2$ are
\begin{equation} \label{eq:cst} g = 2v = \pi J a \; .  \end{equation} It is
now straightforward to obtain the variation of the ground state energy per
site with $m_0$ as \begin{equation} \label{eq:de} E(m_0) -E(m_0=0) = -2
J_\perp m_0^2 - \frac{7}{5 \sqrt[3]{4}} a v n_0^{2/3}
\left(\frac{h}{v}\right)^{4/3} \; , \end{equation} where $n_0$ is the
fermion density, equal to $1/(2a)$ in the original lattice model. Minimizing
with respect to $m_0$ one finds immediately the equilibrium value of the
staggered magnetization as \begin{equation} \label{eq:m0} m_0 =
\frac{28}{15} \left(\frac{14}{15\pi}\right)^{1/2}
\left(\frac{|J_\perp|}{J}\right)^{1/2} \approx 1.017
\left(\frac{|J_\perp|}{J}\right)^{1/2} \; .  \end{equation} The mass gap
then is \begin{equation} \label{eq:gap} \Delta = \frac{56\sqrt{3}}{5 \pi}
|J_\perp| \approx 6.175 |J_\perp| \; .  \end{equation} The results
(\ref{eq:m0}) and (\ref{eq:gap}) will be compared to experiment below.

We now turn to the spin dynamics. The appropriate generalization of the
above mean--field approximation then is an RPA treatment of the interchain
interaction. In the ordered phase translational symmetry is broken, and
therefore there are umklapp processes coupling modes at wavevectors
$\bbox{q}$ and $\bbox{q}+\bbox{Q}$ (where $\bbox{Q}=(0,0,\pi/a)$). The
transverse susceptibility then is a $2\times 2$ matrix given by
\begin{equation} \label{eq:ch3}
\chi(\bbox{q},\omega) = \frac{\chi(q_z,\omega)}{1- 2 |J_\perp| (\cos q_x +
\cos q_y) \chi(q_z,\omega)} \; , 
\end{equation}
where 
\begin{equation} \label{eq:ch1}
\chi(q_z,\omega) = \left( 
\begin{array}{cc}
\chi_n (q_z,\omega)& \chi_u(q_z,\omega) \\
\chi_u(q_z,\omega) & \chi_n(q_z+\pi,\omega) 
\end{array} \right)
\end{equation}
is the susceptibility matrix of the one--dimensional model $H_1$.  Explicit
expressions for $\chi_{n,u}$ are not known, however, a great deal can be
learned from general symmetry properties of $H_1$ and $H_2$: (i) by spin
rotational invariance the magnitude of the staggered magnetization is
independent of the orientation of the staggered field in $H_1$; this implies
$\chi_n(\pi,0) = -1/(4J_\perp)$ and guarantees the existence of a Goldstone
mode (spin wave) in the ordered state. (ii) From the equation of motions
derived from $H_1$ one finds $ \chi_n(0,\omega)=
(h^2/\omega^2)[\chi_n(\pi,\omega)-\chi_n(\pi,0)]$, $ \chi_u(0,\omega) =
\chi_u(\pi,\omega) = (h/\omega)[\chi_n(\pi,\omega)-\chi_n(\pi,0)]$, at $q_z
= 0, \pi$ everything is thus determined by $\chi_n(\pi,\omega)$ alone. (iii)
when all relevant energies ($\omega,\Delta,...$) are much smaller than $J$
the relativistic invariance of $H_2$ can be used: the operators $S^+(q_z
\approx \pi)$ and $S^+(q_z \approx 0)$ are a Lorentz scalar and vector,
respectively. Consequently, $\chi_n(\pi+q,\omega)$ is a function of
$\omega^2-v^2q^2$ only, and $ \chi_u(q,\omega) =
(h\omega/(\omega^2-v^2q^2))[\chi_n(\pi+q,\omega)-\chi_n(\pi,0)]$.  The full
three--dimensional $\chi(\bbox{q},\omega)$ thus is entirely determined by
$\chi_n(\pi+q,\omega)$ alone, as long as $q_z$ is in the vicinity of $0$ or
$\pi$ and $\omega \ll J$.

We can further use the known spectrum of $H_2$ \cite{bergknoff_mtm_exact} to
determine the form of $\chi_n(\pi+q,\omega)$: this function involves
intermediate states where the $z$--component of the magnetization has
increased by $\Delta S^z = 1$, corresponding to an added fermion in the
Thirring model language. These excitation have energy $\Delta$, leading to a
pole. There are of course also multiple--excitation contributions, leading
to a continuous spectrum. The lowest such excitation comes from a
combination of the elementary $\Delta S^z = 1$ excitation with an $\Delta
S^z = 0$ excitation which also has energy $\Delta$ (see the discussion of
the longitudinal excitations below), thus leading to a threshold at
$\omega=2\Delta $. Notice that the elementary excitations creating this
continuum are thus quite different from the spinons responsible for the
lowest continuum of an isolated chain.\cite{bethe_xxx} Multiple--excitation
continua of course also exist, with thresholds $n\Delta, n\ge 3$. One thus
can write
\begin{equation} \label{eq:chin}
\chi_n(\pi+q,\omega) = \frac{z}{\Delta^2 + v^2q^2-\omega^2} +
f(\omega^2-v^2q^2) \; ,
\end{equation}
where the unknown function $f(x)$ contains contributions from the
many--excitation continua, has a threshold singularity at $x=4\Delta^2$,
and is real below the threshold. The constant $z$ ensures that $\chi_n(\pi,0) =
-1/(4J_\perp)$. 

The excitation spectrum now is given by the singularities of
$\chi(\bbox{q},\omega)$, eq.(\ref{eq:ch3}), and in particular the
low--energy ($\omega<2\Delta$) states are found from poles. One finds: (i)
for propagation along the chain there is a spin wave mode with
$\omega(0,0,q_z) = v(J_\perp)|q_z|$, where the the spin wave velocity is
only weakly affected by interchain coupling: e.g. neglecting $f$ in
eq.(\ref{eq:chin}) (the ``single mode approximation'', SMA), $v(J_\perp) =
v/\sqrt{1+h^2/\Delta^2}$; (ii) for transverse wavevector $(\pi,0)$ and more
generally on the whole line $\cos q_x + \cos q_y =0$ the spin wave frequency
is entirely determined by the one--dimensional result, and therefore from
eq.(\ref{eq:chin}) $\omega(\pi,0,\pi) = \Delta$, i.e. {the mass gap is
directly accessible experimentally and proportional to $J_\perp$}; (iii) in
the SMA the spin--wave dispersion is $\omega(\bbox{q})^2= \Delta^2 (1-(\cos
q_x + \cos q_y)/2) + v^2q_z^2$. In the relevant case $|J_\perp| \ll J$ the
$\bbox{q}$--dependence of the transverse part that of standard spin wave
theory.  Taking into account $f$ will lead to a modified relation in the
transverse directions; (iv) in the SMA, the static susceptibility is
$\chi(0,0) = 14 \pi/(405 J) \approx 0.109/J$. Taking into account $f$ will
decrease this value. One can notice that for $J_\perp \rightarrow 0$ one
expects to recover the purely one--dimensional result $\chi(0,0) =
1/(\pi^2J) \approx 0.101/J$. This suggests that the SMA is a rather good
description of low--energy properties.

There also are longitudinal excitations, corresponding to oscillations of
$m_0$ about its mean value. Within the RPA, the longitudinal susceptibility
$\chi_L$ is given by a formula analogous to eq.(\ref{eq:ch3}). From
eq.(\ref{eq:de}) $\chi_L(\pi,0) = -1/(12 J_\perp)$, and the frequency
dependence can again be obtained from the excitation spectrum of the massive
Thirring model: here excitations with $\Delta S^z = 0$ intervene,
i.e. particle--hole pairs. Naively, one expects a continuum above a gap
$2\Delta$, however, because of the interaction in $H_2$, the lowest
excitation is actually an excitonic bound state at energy
$\Delta$,\cite{bergknoff_mtm_exact} i.e. at the same energy as the lowest
$\Delta S^z = \pm 1$ excitation.\cite{rem2} The one--dimensional $\chi_L$
then has the same form as $\chi_n$, eq.(\ref{eq:chin}), and in the SMA,
there then is a longitudinal mode at $\omega_L^2(\bbox{q}) = \Delta^2[1
-(\cos q_x + \cos q_y)/6] + v^2 q_z^2$.  Going beyond the SMA, the already
relatively weak dispersion in the transverse directions would be further
reduced. Notice that at $\bbox{q}_\perp = (\pi,0)$ longitudinal and transverse
modes are degenerate.

In this approximation, the longitudinal mode is well--defined. This is
however an oversimplification: a longitudinal mode can decay into two spin
waves.\cite{rem1} To get a quantitative estimate of this effect I have
obtained an effective Ginzburg--Landau description of the model via a
Hubbard--Stratonovich transformation of the interchain interaction in
eq.(\ref{eq:h0}). At the Gaussian level expressions like eq.(\ref{eq:ch1})
for the transverse and longitudinal susceptibilities are obtained, but
higher order corrections can also be studied systematically. In particular,
a longitudinal excitation can now decay into two spinwaves, and there is
also important spinwave--spinwave scattering. Spin rotation invariance
imposes that the matrix elements for these processes at low energies are
given by $\chi_L(\pi,0)/h$ and $\chi_L(\pi,0)/h^2$, respectively. Taking
these processes into account in an RPA--like fashion, the longitudinal
susceptibility takes the form
\begin{equation} \label{eq:}
\chi_L(\bbox{q},\omega) = \frac{\chi_L(q_z,\omega) +
\Sigma(\bbox{q},\omega)}%
{1-2|J_\perp|(\cos q_x + \cos q_y) (\chi_L(q_z,\omega) +
\Sigma(\bbox{q},\omega))} \; .
\end{equation}
This form still retains one--dimensional relativistic invariance:
$\chi_L((\bbox{q}_\perp, q_z),\omega) =\chi_L((\bbox{q}_\perp,
0),(\omega^2-v^2 q_z^2)^{1/2})$.  Numerical results for ${\rm Im}[\chi_L]$
(which determines the neutron scattering intensity) are shown in
fig.\ref{f1}. One notices in particular a very sharp feature very close to
the pole of the unrenormalized ($\Sigma =0$) propagator. Numerically, the
width of this peak is found of order $0.01 \Delta$, the transverse mode
remains thus rather well defined even when decay into spinwaves is taken
into account. This can be attributed to the low frequency of the mode and
the limited phase space available for decay into spinwaves. In addition to
the peak, one also notices an incoherent background which for $q_z=\pi$
extends down to zero energy.

Finally, the N\'eel temperature can be determined in this approach from the
divergence of the static susceptibility at $\bbox{q}=(0,0,\pi)$ as a
function of temperature.  The single--chain susceptibility is given by
\cite{giamarchi_logs,affleck_log_corr} $\chi(\pi,0;T)= A/(JT)
\ln^{1/2}(\Lambda J/T)$, where numerical calculations give \cite{rem3} $A
\approx 0.32, \Lambda \approx 5.8$. Eq.(\ref{eq:ch3}) then implies the
relation 
\begin{equation} \label{eq:tn} 
|J_\perp| = \frac{T_N}{4A
\ln^{1/2}(\Lambda J/T_N)} \; .  
\end{equation}

The above results for staggered magnetization, excitation spectrum, and
N\'eel temperature are now compared to experimental results on $\rm KCuF_3$.
From the excitation spectrum above $T_N$ the exchange constant along the
chains is $J=34{\rm meV}$. From the measured \cite{satija_kcuf} $m_0 =0.25$
eq.(\ref{eq:m0}) then gives $J_\perp = -J/16 =-2.1{\rm meV}$. From
eq.(\ref{eq:gap}) then follows the first prediction of the present theory:
$\Delta = 13{\rm meV}$. From eqs.(\ref{eq:ch3}) and eq.(\ref{eq:ch1})
$\Delta$ is the spin wave frequency at
$\bbox{q}_1=(\pi,0,\pi)$. Experimentally \cite{satija_kcuf}
$\omega(\bbox{q}_1) =11.5{\rm meV}$, in rather good agreement with the
prediction. A second nontrivial prediction of the present theory is the
existence of a longitudinal mode, at $\omega_L(0,0,\pi) \approx \sqrt{2/3}
\Delta \approx 10{\rm meV}$. It is then tempting to associate the sharp rise
in the energy--dependent neutron scattering intensity (which does not
differentiate between transverse and longitudinal modes) observed around
$\omega=10 {\rm meV}$ \cite{satija_kcuf,tennant_kcuf_q1d} with this mode. On
the other hand, it is not clear how the two--spinwave process proposed in
ref.\cite{tennant_kcuf_q1d}, which is rather featureless around
$\omega=\Delta$ (see fig.\ref{f1}), can account for this result. It would
clearly be of interest to study this point in more detail.

Using the estimated values for $J$, $J_\perp$ eq.(\ref{eq:tn}) gives an
estimate for the N\'eel temperature: $T_N \approx 60K$, overestimating the
experimental result ($T_N=39K$) by about 50\%. This discrepancy is in part
due to the fact that logarithmic correction terms, enhancing the tendency to
ordering, are included in eq.(\ref{eq:tn}), but neglected in the initial
fermionic model, eq.(\ref{eq:h2}). In the fermionic language, the
logarithmic terms come from an extra umklapp interaction
\cite{haldane_xxzchain} $g_u(\psi_R^\dagger
\psi_R^\dagger\psi_L\psi_L+h.c.)$ in eq.(\ref{eq:h2}). These terms destroy
the solvability of the fermionic model, but can be taken into account
perturbatively. \cite{affleck_log_corr} To lowest order then the $h^{4/3}$
term in eq.(\ref{eq:de}) is multiplied by a factor $(1 + y_0 \ln
(v/\Delta))^{1/3}$, with $y_0 = g_u/(\pi v)$. With $y_0 \approx 0.25$
\cite{affleck_log_corr} the experimental value of $m_0$ leads to $J_\perp =
-1.6 \rm meV$. From this $\Delta = 11.4 \rm meV$, very close to the
experimental value, and $T_N = 47 K$. The remaining discrepancy between
theory and experimental values of $T_N$ may well be related to thermal
fluctuation effects neglected here and which can be estimated to be on the
level of 10\%.\cite{schulz_gl_q1d} Experiment indeed indicates persistence
of short--range order well above $T_N$.\cite{satija_kcuf,tennant_kcuf_q1d}

It is also interesting to consider the case of $\rm
Sr_2CuO_3$.\cite{keren_srcuo,ami_srcuo} From the experimental $T_N \approx
5K$ and the estimated $J \approx 220 \rm meV$ one gets $J_\perp \approx 0.12
\rm meV$. This then leads to a predicted spin wave energy at $\bbox{q} =
(\pi,0,\pi)$ of $\Delta = 1.1 \rm meV$ and $m_0 = 0.036$. This last number
is consistent with the experimental upper limit $m_0 \leq
0.05$.\cite{ami_srcuo} It should however be noticed that in $\rm Sr_2CuO_3$
there is considerable structural anisotropy in the directions perpendicular
to the chains, and the value derived for $J_\perp$ therefore is an average
over transverse directions.

The continuum limit used in the present treatment is valid if the
correlation length $\xi = v/\Delta \approx v/(\pi T_N)$ is large compared to
the lattice constant. For the two compounds discussed above one finds $\xi =
5a$ and $\xi = 320 a$, satisfying this criterion. The mean--field treatment
of the interchain coupling leads to $T_N \propto J_\perp$, $m_0 \propto
\sqrt{|J_\perp|/J}$, up to logarithmic corrections. These relations are
consistent with general scaling arguments.\cite{grower_cross} On the other
hand, the frequently used renormalized spin wave theory \cite{oguchi_spinwa}
predicts $T_N \propto \sqrt{J |J_\perp |}$, $m_0 \propto
1/\ln(J/|J_\perp|)$, in clear contradiction with scaling. Apart from the
quantitatively satisfying description of $\rm KCuF_3$, the present theory
thus also seems more satisfactory on grounds of consistency with generally
valid arguments.

I am grateful to A. Keren for interesting comments and to D. A. Tennant and
A. Sandvik for stimulating correspondence. Laboratoire de Physique des
Solides is a Laboratoire Associ\'e au CNRS.

\begin{figure}
\caption[t]{Frequency dependence of the imaginary part of the longitudinal
susceptibility (in units of $1/|J_\perp|$) at $\bbox{q}=(0,0,\pi+q_z)$ for
$q_z=0$ (full line), $vq_z=\Delta /2$ (dashed), $vq_z=\Delta$
(dash--dotted), and $vq_z=2\Delta$ (dash--double--dot). Also shown is the
two--spinwave cross section at $q_z=0$, according to
ref.\cite{tennant_kcuf_q1d} (double--dash--dot).}
\label{f1}
\end{figure}

%\bibliography{revues,1dtheory,1dexp,nchains_1}

\begin{thebibliography}{10}
\bibitem{rem2}
This equality only holds for small interchain coupling (small $h$). For large
  $h$ it is clear from eq.(\ref{eq:h1}) that $\Delta_{\Delta S^z = 0} = 2
  \Delta_{\Delta S^z = 1} = 2h$.
\end{thebibliography}
%\bibliographystyle{prsty}

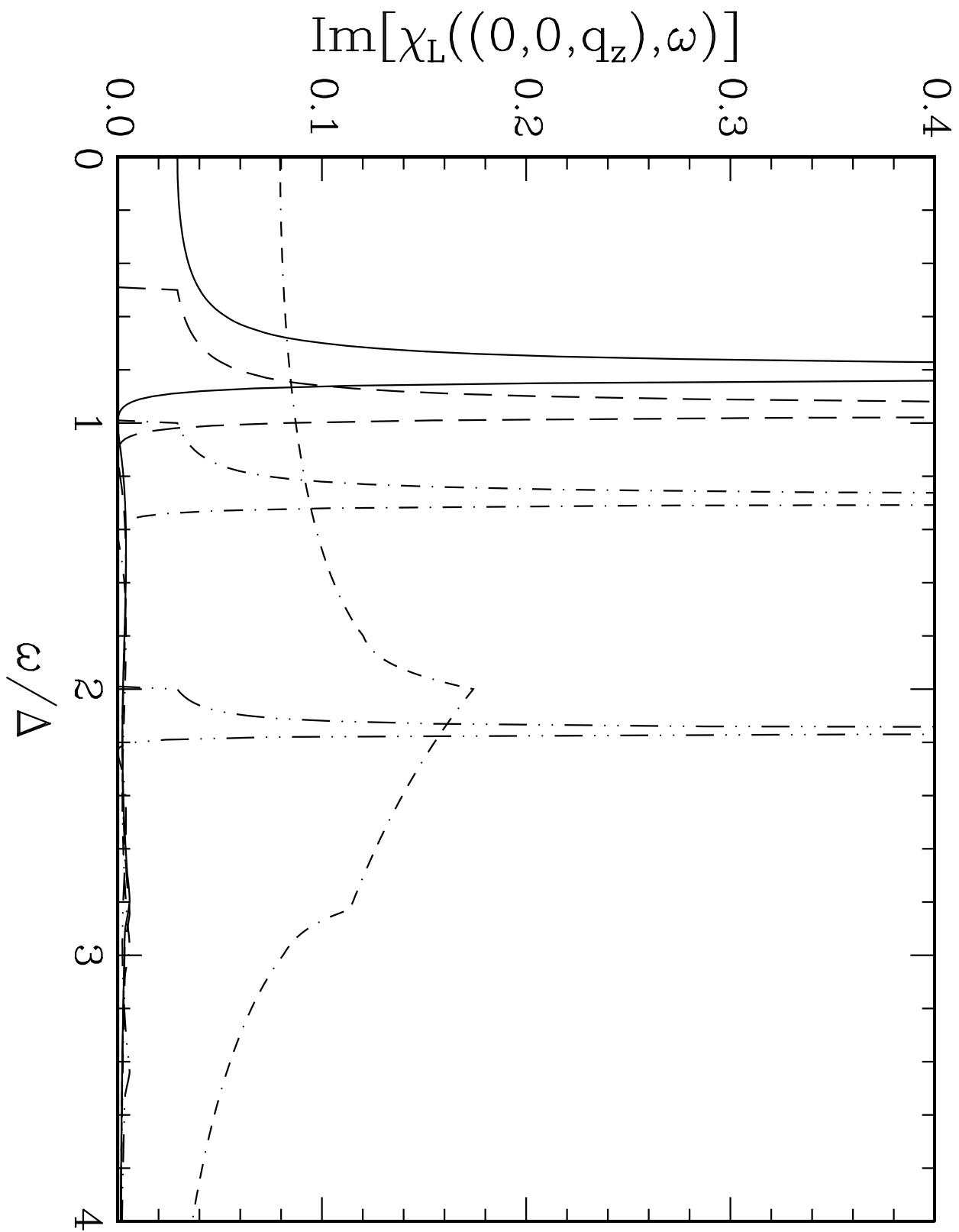
\end{document}